\documentclass[conference]{IEEEtran}
\usepackage[hyphens]{url}
\usepackage{hyperref}
\usepackage[backend=biber, style=ieee]{biblatex}

\usepackage{amsmath,amssymb,amsfonts}
\usepackage{algorithmic}
\usepackage{graphicx}
\usepackage{textcomp}
\usepackage{xcolor}
\usepackage{balance}
\usepackage{booktabs}
\usepackage{tikz}
\usepackage{array}
\usepackage{float}
\usepackage{lipsum}

\graphicspath{{./images/}}

\usepackage{balance}

\begin{document}

\date{}

\title{A Survey of Unikernel Security: Insights and Trends from a Quantitative Analysis}

\author{\IEEEauthorblockN{Alex Wollman}
\IEEEauthorblockA{\textit{The Beacom College of Computer and Cyber Sciences} \\
\textit{Dakota State University}\\
Madison, SD, USA \\
0009-0000-6260-2750}
\and
\IEEEauthorblockN{John Hastings}
\IEEEauthorblockA{\textit{The Beacom College of Computer and Cyber Sciences} \\
\textit{Dakota State University}\\
Madison, SD, USA \\
0000-0003-0871-3622}
}

\maketitle

\begin{abstract} \label{Abstract}
Unikernels, an evolution of LibOSs, are emerging as a virtualization technology to rival those currently used by cloud providers. Unikernels combine the user and kernel space into one ``uni''fied memory space and omit functionality that is not necessary for its application to run, thus drastically reducing the required resources. The removed functionality however is far-reaching and includes components that have become common security technologies such as Address Space Layout Randomization (ASLR), Data Execution Prevention (DEP), and Non-executable bits (NX bits). This raises questions about the real-world security of unikernels. This research presents a quantitative methodology using TF-IDF to analyze the focus of security discussions within unikernel research literature. Based on a corpus of 33 unikernel-related papers spanning 2013-2023, our analysis found that Memory Protection Extensions and Data Execution Prevention were the least frequently occurring topics, while SGX was the most frequent topic. The findings quantify priorities and assumptions in unikernel security research, bringing to light potential risks from underexplored attack surfaces. The quantitative approach is broadly applicable for revealing trends and gaps in niche security domains.
\end{abstract}


\section{Introduction} \label{Introduction}

Virtualization of compute environments began as early as the 1960s as a means to enhance mainframe performance \cite{VirtConTax}. Continuing through the late 20th and beginning of the 21st century, virtualization saw many advancements including virtual machines (VMs) and containers \cite{CompVirtandCont,Virt_IssuesAndSolutions}. These advancements provided both better performance and resource utilization \cite{CompVirtandCont,VMSurvey,SecuritySurvey}, but for a growing cloud environment more advancements were needed \cite{happe2017unikernels,InflateUnikernels}.

Unikernels are an evolution of an old idea: Library Operating Systems (LibOS). The origins of LibOSs stem from the early 1990s \cite{RethinkingLibOS}. The principle idea behind LibOSs is that OS functionality is implemented in the form of libraries, customized to increase performance of the application by providing closer access to hardware \cite{RethinkingLibOS}. However, the rise of Virtual Machines (VMs) around the same time quickly overshadowed LibOSs \cite{Disco} as a virtualization technique. VM's ability to replicate the entire OS and run applications, compared to the customization required for LibOSs \cite{RethinkingLibOS}, put LibOSs at a disadvantage. Additionally, VMs provided a way for several OSs to run on one physical device \cite{Virt_IssuesAndSolutions}. Several decades later, after the flourishing of technological advances like the Internet, LibOSs reimerged as unikernels. \cite{Unikernel:LibOSForTheCloud}.

Unikernels are different from VMs and containers in several key ways. Unikernels, as the name implies, ``uni''fy the user and kernel address spaces into one \cite{Unikernel:LibOSForTheCloud,UnikernelApproachCloud}. This also means that when the unikernel is compiled, all necessary kernel functionality along with the application code are combined into one binary \cite{Unikernel:LibOSForTheCloud,FreshLookAtIsolationPlatforms}. Combining the application and kernel into one binary comes with some benefits. Utilizing syscalls to invoke kernel functionality and context switches into and out of kernel are no longer required, and simpler function call implementations can be used instead \cite{Unikernel:LibOSForTheCloud, RethinkingLibOS, unikernelIsolation_MPK}. The direct invocation of kernel code in this way is where the tradition of LibOSs shines through.

Furthermore, unikernels have a drastically smaller memory requirement than other virtualization techniques \cite{Unikernel:LibOSForTheCloud, FreshLookAtIsolationPlatforms}. Unikernels achieve this through a build system specifically designed to compile in only the necessary components required to execute the unikernel's application \cite{UnikernelApproachCloud, Unikernel:LibOSForTheCloud}. In contrast, if a Linux VM is built to host a web server, that VM is built with all the core Linux utilities and libraries in addition to any that are required by the web server. Even containers, while not comprising the entire OS, still possess a plethora of functionality above and beyond its intended function.

To help achieve the reduced size, many familiar security features have also been removed including Address Space Layout Randomization (ASLR), Non-executable Bits (NX bits), and stack canaries \cite{michaels2019assessing, Unikernel:LibOSForTheCloud, UnikernelSecurityPerspective}. Adhering to the reductionist principle adopted by unikernels, these features, in addition to others, have been removed to save space, simplify the code base, and eliminate extra, unnecessary features \cite{Unikernel:LibOSForTheCloud, michaels2019assessing, UnikernelSecurityPerspective}. The majority of unikernel literature refer to this reduction of capabilities as a security benefit \cite{CloudCyberSecurity, UnikernelSecurityPerspective}, but not all agree \cite{michaels2019assessing}. Similar motivations, if applied to the development cycle of a prominent OS such as Windows, OSX, or Linux, combined with 20 years of advances in exploitation techniques would inevitably lead to a drastic increase in exploit development research. Is this happening in unikernel research?

To conduct a deeper investigation into the implications of removing such security components, the goal of this research is to identify trends in security research related to unikernels. Specifically, through a quantitative analysis of security terms and unikernel names in unikernel research literature, this study seeks to answer the following research questions:
\begin{itemize}
    \item \textbf{RQ1}: What security term appears most frequently in the corpus?
    \item \textbf{RQ2}: What security term appears least frequently in the corpus?
    \item \textbf{RQ3}: What is the publication trend for unikernel related research?
    \item \textbf{RQ4}: Does the frequency analysis provide any insight into unikernel security?
\end{itemize}

The outcome of this study is intended to provide insight into unikernel security trends and to help inform future research efforts. \textbf{RQ2} specifically intends to provide data to demonstrate what additional research into unikernel security is needed. \textbf{RQ3} and \textbf{RQ4} intend to investigate publication trends and show if additional research is needed into unikernels.

The rest of this paper is organized as follows. Section \ref{StudyDesign} addresses the literature review, its methodology, and the quantitative analysis methodology. Section \ref{Results} shares the findings of the research. Section \ref{Discussion} offers a discussion and insight into the findings of the research. Section \ref{RelatedWork} discusses related work, and Section \ref{FutureWork} discusses future research avenues. Finally, Section \ref{Conclusion} presents the conclusion reached and through this study.

\section{Study Design} \label{StudyDesign}

\subsection{Literature Review Methodology} \label{LitRevMethod}

To answer the research questions specified in Section \ref{Introduction}, the following research goals are established:
\begin{itemize}
    \item \textbf{RG1}: Collect a diverse corpus of research literature related to unikernels.
    \item \textbf{RG2}: Calculate the frequency of security terms and unikernel names in literature.
    \item \textbf{RG3}: Determine the significance of the terms in relation to the corpus of literature.
\end{itemize}

To satisfy \textbf{RG1}, a literature review methodology was developed to ensure only unikernel related literature would be included in this study. The methodology follows three primary steps detailed in the following sections: \ref{LitSearch} Literature Search, \ref{LitSelection} Literature Selection, and \ref{FreqCalc} Data Extraction.

\subsection{Literature Search} \label{LitSearch}
In the literature search a tailored, but generic, list of search terms was used to generate an initial set of literature. 
In order to conduct a thorough investigation into unikernel security, a corpus of literature was collected utilizing different databases.

\subsubsection{Database Selection} \label{LitDatabases}
This study primarily utilized three databases: (1) IEEE Xplore, (2) The ACM Digital Library, and
(3) Google Scholar. IEEE Explore and The ACM Digital Library were selected based upon size, popularity, relevance to the
subject, and reputation. Google Scholar was selected to broaden the search scope and utilize alternative data sources, such as Springer.

\subsubsection{Search Method}
Search terms such as ``unikernels'', ``unikernel security'', and ``unikernel debugging'' were supplied to the databases listed in \ref{LitDatabases} to generate initial results. These results were read by the researcher and used to create additional, more specific, search terms such as ``MirageOS vulnerabilities'', ``SGX in unikernels'', ``memory corruption in unikernels'', and ``security overview of unikernels''.

\subsection{Literature Selection} \label{LitSelection}
The literature selection phase utilized exclusion and inclusion criteria to generate a relevant corpus of literature. From the search results, each research paper was evaluated using a simple check list:
\begin{itemize}
    \item Is the paper about unikernels?
    \item Does the paper involve unikernels in a substantial manner?
\end{itemize}

The simplicity of this list is meant to gather a variety of topics in order to better evaluate the field as a whole. The literature was neither screened in favor of, nor against, particular topics in order to avoid artificially impacting the Term Frequency - Inverse Document Frequency (TF-IDF) \cite{DocumentClustering, textMiningTF_IDF} calculation described in \ref{FreqCalc}. For instance, only selecting security related research papers could skew the TF-IDF value showing a greater than expected frequency of security terms. Similarly, selecting only development focused research papers could have the opposite effect. It was also important to ensure that research papers had more than a passing reference to unikernels. Referencing a specific unikernel once in a research paper could give the impression it is exceedingly rare.

\subsection{Data Extraction and Frequency Calculation} \label{FreqCalc}
To address \textbf{RG2} and \textbf{RG3}, data is extracted from the research papers to calculate how frequently different security terms are used in relation to unikernels. During the data extraction phase, literature is scanned for specific security terms and unikernel names. A security term list was used as the basis to scan the literature and extract the data used to calculate the TF-IDF values. This list also provided a standard way to extract and eventually compare the collected literature.

The security terms were selected using two methods. The first method comprised the manual extraction of terms from the corpus of literature. Each research paper was read and every unique security term added to a list, including common acronyms and abbreviations. Accounting for these acronyms is important because the full term is often only used once, or only at the beginning of research papers or sections, and its acronym used elsewhere. The second method utilized the researchers' experience, including terms that may not appear in the corpus but are still relevant to the security field. A sample of security acronyms is shown in Table \ref{table_SecurityAcronyms}.

\begin{table}[ht]
    \centering
    \caption{Security Acronyms}
      \begin{tabular}{|c|}
          \hline
          Sample Security Acronyms\\
          \hline
          ASLR, MPK, NX-Bit, CPI\\
          \hline
      \end{tabular}
  \label{table_SecurityAcronyms}
  \end{table}

The search term list also included unikernel names gathered using the same methods described for security terms. To account for variations in term appearance, a term is included in the list multiple times utilizing regular expressions (regex.) The most common variation utilized a single dash (``-'') between words: software-defined security versus software defined security. The full list of both security and unikernel terms are provided Tables \ref{table_fullSecurityTerms} and \ref{table_unikernelTermList}, respectively.

\begin{table}[h]
\centering
\caption{Security Search Term List}
    \begin{tabular}{|c|c|}
        \hline
        ASLR & code-pointer\textbackslash s?integrity\\
        cfi & address\textbackslash s?space\textbackslash s?layout\textbackslash s?randomization\\
        cpi & control\textbackslash s?flow\textbackslash s?integrity\\
        dep & control-flow\textbackslash s?integrity\\
        mpk & data\textbackslash s?execution\textbackslash s?prevention\\
        mpx & memory\textbackslash s?protection\textbackslash s?keys\\
        NX-bit & memory\textbackslash s?protection\textbackslash s?extensions\\
        NX\textbackslash s?bit & return\textbackslash s?oriented\textbackslash s?programming\\
        rop & non-executable\textbackslash s?bit\\
        sgx & software-fault\textbackslash s?isolation\\
        sfi & software\textbackslash s?fault\textbackslash s?isolation\\
        sdsec & software\textbackslash s?guard\textbackslash s?extensions\\
        no-executable\textbackslash s?bit & software-defined\textbackslash s?security\\
        stack\textbackslash s?canaries & software\textbackslash s?defined\textbackslash s?security\\
        \hline
    \end{tabular}
\label{table_fullSecurityTerms}
\end{table}

\begin{table}[h]
\centering
\caption{Unikernel Search Term List}
    \begin{tabular}{|c|c|}
        \hline
        hermitux & unik \\
        minios & clive \\
        halvm & graphene-sgx \\
        clickos & azalea \\
        mirageos & drawbridge \\
        includeos & ling \\
        nanos & guk \\
        osv & runtime.js\\
        occlum & unikraft\\
        rumprun & torokernel\\
        \hline
    \end{tabular}
\label{table_unikernelTermList}
\end{table}

To determine the significance of security terms, and address \textbf{RG2}, the Term Frequency - Inverse Document Frequency (TF-IDF) algorithm is used. Term Frequency (TF) is a mathematical equation used to determine, per document, the frequency of a term. 

TF is defined as the total number of times the specific term $t$ in document $d$ occurs divided by the summation of all terms $t\prime$ in the same document $d$ \cite{textMiningTF_IDF}. The TF value can be useful for comparison of terms within a document, but is not as useful for comparing between documents due to variations in document length. For this research a normalized method was utilized to help account for variations in document length, resulting in an addition and multiplication of 0.5 as shown in the mathematical formula below.

\[TFreq(t,d) = 0.5 + 0.5(\frac{f_{t,d}}{\sum_{t\prime\in d}})\]

Inverse Document Frequency (IDF) assigns a weight to each word in a document, inversely proportional to how often the term occurs in a corpus of documents: the more frequent the term the lower the weight \cite{DocumentClustering, textMiningTF_IDF}. IDF is defined as the logarithm\footnote{When researching this algorithm there was ambiguity in the sources for which logarithm to use: base 10 or natural log. For this research the base 10 logarithm was used.} of the total number of documents $N$ divided by the number of documents $d$ within the overall corpus $D$ in which the term $t$ appears \cite{textMiningTF_IDF}.
To avoid a possible divide by zero error if a term does not appear, 1.0 is added to the numerator and denominator, as shown in the mathematical formula below.

\[IDFreq(t,d)= \log(\frac{N+1.0}{(d\in D:t\in d)+1.0})\]

The TF-IDF is then calculated by multiplying the TF and IDF values. The more documents in which the term appears brings the TF-IDF value closer to 0.0 \cite{DocumentClustering, textMiningTF_IDF}. Therefore, TF-IDF values that are closer to 1.0 indicate terms that appear less frequently in the corpus. While the formula is straight forward, it is provided below for completeness.

\[f(t,d) = TFreq(t,d) * IDFreq(t,d)\]

This research utilized Python3 version 3.10.12 to write the artifact which scans the corpus of literature and generates the TF, IDF, and TF-IDF values utilizing the term list as shown in Table \ref{table_fullSecurityTerms}. The Python package PyPDF2 \cite{PyPDF2} reads the PDF documents, and the built-in Python3 regular expression package ``re`` searches for the terms. Each line of the PDF is read and scanned, searching for each term in the list. An internal structure is used to track the number of occurrences, per research paper, and the sentence containing the term. When all documents are processed, the TF, IDF, and TF-IDF values are calculated.

\section{Results} \label{Results}

\subsection{Overview}
A total of 33 papers were selected that fit the criteria laid out in Section \ref{LitRevMethod}. The literature corpus was published between March 2013 and May 2023, with one outlier from 1995. Table \ref{PaperCounts} shows the number of papers used and from which database they were discovered. Table \ref{table_corpusPapers} lists the discovered papers. The majority of papers were found through both ACM (14) and IEEE (12), with the remainder through Google Scholar (7). The individual entities comprising the Google Scholar papers were Defense Technical Information Center, NCCGroup, Usenix, Security in Computing and Communications, Network and Distributed System Security, and the University of the Netherlands. An observation while conducting the search, as described in \ref{LitSearch}, is that the current state of unikernel research primarily focuses on development, with some research applying processor specific techniques such as Software Guard Extensions (SGX) \cite{Uniguard} and Memory Protection Keys \cite{unikernelIsolation_MPK}.

\begin{table}[H]
\centering
\caption{Paper Sources and Counts}
\begin{tabular}{|c|c|}
    \hline
    Paper Source & Paper Count\\
    \hline
    ACM & 14\\
    Google Scholar & 7\\
    IEEE & 12\\
    \hline
\end{tabular}
\label{PaperCounts}
\end{table}

\begin{table}[ht!]
  \centering
  \caption{Corpus of Papers}
        \begin{tabular}{|m{30em}|}  
            \hline
            Paper Names\\
            \hline
            SCONE: Secure Linux Containers with Intel SGX\cite{SCONE}\\ 
            OS Noise Analysis on Azalea-Unikernel\cite{OSNoiseAzalea}\\
            Demo: On-The-Fly Generation of Unikernels for Software-Defined Security in Cloud Infrastructures\cite{OnTheFlyUnikernel}\\
            A TOSCA-Oriented Software-Defined Security Approach for Unikernel-Based Protection Clouds\cite{TOSCABasedUnikernelApproach}\\
            Unikernel-based Approach for Software-Defined Security in Cloud Infrastructures\cite{UnikernelApproachCloud}\\
            A Survey on Security Isolation of Virtualization, Containers, and Unikernels\cite{SecuritySurvey}\\
            Cloud Cyber Security: Finding an Effective Approach with Unikernels\cite{CloudCyberSecurity}\\
            Exokernel: An Operating System Architecture for Application-Level Resource Management\cite{ExokernelResMan}\\
            USETL: Unikernels for the Serverless Extract Transform and Load; Why Should You Settle for Less?\cite{USETL:Unikernels}\\
            Want More Unikernels? Inflate Them!\cite{InflateUnikernels}\\
            Understanding and Hardening Linux Containers\cite{grattafiori2016understanding}\\
            MirageOS Unikernel with Network Acceleration for IoT Cloud Environments\cite{MirageOSforIoT}\\
            Azalea Unikernel IO Offload Acceleration\cite{AzaleaAccel}\\
            Unikernel Network Functions: A Journey Beyond the Containers\cite{UnikernelNetFunc}\\
            Unikraft and the Coming Age of Unikernels\cite{lefeuvre2021unikraft}\\
            fASLR: Function-Based ASLR Via TrustZone-M and MPU for Resource-Constrained IoT Systems\cite{fASLR}\\
            Unikernels: Library Operating Systems for the Cloud\cite{Unikernel:LibOSForTheCloud}\\
            NCC Group Assessing Unikernel Security\cite{michaels2019assessing}\\
            A Syscall-Level Binary-Compatible Unikernel\cite{SyscallUnikernel}\\
            Rethinking the Library OS from the Top Down\cite{RethinkingLibOS}\\
            Unikernel Linux (UKL)\cite{UnikernelLinux}\\
            Virtualization: A Survey on Concepts, Taxonomy And Associated Security Issues\cite{VirtConTax}\\
            Virtualization and Containerization of Application Infrastructure: A Comparison\cite{CompVirtandCont}\\
            Uniguard Protecting Unikernels using Intel SGX\cite{Uniguard}\\
            Occlum: Secure and Efficient Multitasking Inside a Single Enclave of Intel SGX\cite{Occlum}\\
            Panoply: Low-TCB Linux Applications with SGX Enclaves\cite{shinde2017panoply}\\
            A Design and Verification Methodology for Secure Isolated Regions\cite{SecureIsolatedRegions}\\
            Container Security: Issues, Challenges, and the Road Ahead\cite{ContainerSecurity}\\
            Intra-unikernel Isolation with Intel Memory Protection Keys\cite{unikernelIsolation_MPK}\\
            A Security Perspective on Unikernels\cite{UnikernelSecurityPerspective}\\
            A Fresh Look at the Architecture and Performance of Contemporary Isolation Platforms\cite{FreshLookAtIsolationPlatforms}\\
            Unikernels as Processes\cite{UniProc}\\
            Accelerating Disaggregated Data Centers Using Unikernel\cite{AccelDisaggUnikernel}\\
            \hline
        \end{tabular}
        \label{table_corpusPapers}
    \end{table}

\begin{figure}
    \centering
    \includegraphics[scale=.25]{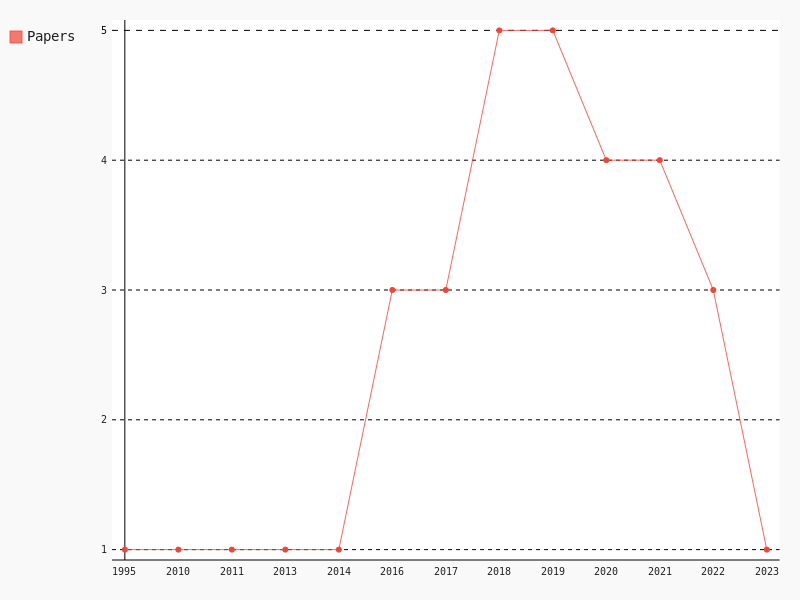}
    \caption{Corpus Published Dates}
    \label{fig:corpus_chart}
\end{figure}

To address \textbf{RQ3}, Figure \ref{fig:corpus_chart} shows how many papers in the corpus were published, per year, between 1995 and 2023. Beginning in 2014 the number of papers steadily increased to five a year in both 2018 and 2019. An interesting observation is the dip in papers beginning in 2019 and continuing through 2023, declining back to one. The literature review methodology as discussed in Section \ref{LitRevMethod} does not restrict publication dates, which indicates this dip is inherent in the data. A possible explanation for this dip could be the COVID-19 pandemic, however an in-depth investigation is outside the scope of this paper.

While collecting papers for the research corpus, only \cite{michaels2019assessing} was found to provide a comprehensive analysis of unikernel security. Its findings showcase vulnerabilities that existing security techniques are designed to fix \cite{michaels2019assessing}, and their removal in unikernels could indicate these vulnerabilities may pose a significant problem once again. 

The development and construction of unikernels is the primary focus of the unikernel literature, with some focusing on processor specific security implementations \cite{Uniguard, unikernelIsolation_MPK} and others discussing security more broadly \cite{CloudCyberSecurity, UnikernelSecurityPerspective}. Some research papers do not refer to the lack of security features as an issue, in part due to removed features such as ``bash'' or ``sh'' which on traditional systems enabled further exploitation \cite{UnikernelSecurityPerspective, michaels2019assessing}. Instead, literature in the corpus relies on the nature of unikernels and their lack of a diversified code base \cite{michaels2019assessing} to provide security, using phrases such as:
\begin{itemize}
    \item ``...unikernel images that integrate protection mechanisms...allow significant reduction of the attack surface.''\cite{UnikernelApproachCloud}
    \item ``Numerous Unikernels do not implement a shell natively, making most types of payloads, that typically rely on bash, ineffective.'' \cite{UnikernelSecurityPerspective}
    \item ``Pivot attacks that depend on shell-access are thwarted.'' \cite{happe2017unikernels}
\end{itemize}

\subsection{TF-IDF Results}
\begin{figure}
    \centering
    \includegraphics[scale=.25]{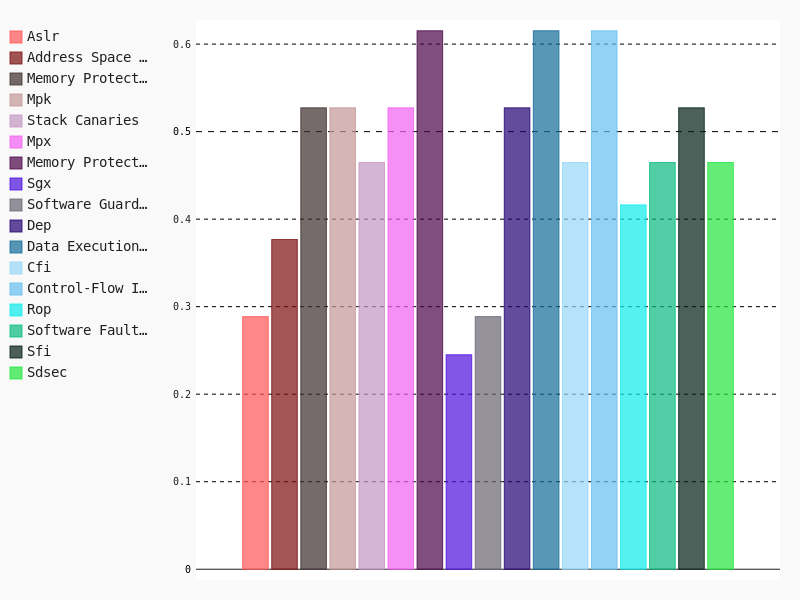}
    \caption{Security TF-IDF Values}
    \label{fig:security_TFIDF}
\end{figure}

The resulting security term TF-IDF values can be seen in Figure \ref{fig:security_TFIDF}. The security terms that appear least frequently with a TF-IDF value of 0.5 or greater are MPK (Memory Protection Keys), Memory Protection Extensions (MPX), Data Execution Prevention (DEP), and Control-Flow Integrity (CFI). These values can be seen in Table \ref{TFIDF_Security_Values} and are denoted by a single star (*).

Addressing \textbf{RQ2}, the least frequently occurring term is tied amongst Memory Protection Extensions, Data Execution Prevention, and Control-Flow Integrity with a value of 0.615. Using their acronyms to break the tie, only CFI is eliminated. Careful observation of the results reveals that a few terms are conspicuously absent: Return Oriented Programming and Software Defined Security. Whereas these terms' acronyms occurred in the corpus, neither of these terms did themselves.

\begin{table}[htbp]
    \centering
    \caption{Security TF-IDF Values}
    \begin{tabular}{|l|c|c|}
        \hline
        Term & TF-IDF Value\\
        \hline
        ASLR** & 0.288\\
        Address Space Layout Randomization & 0.376\\
        CFI & 0.464\\
        Control-Flow Integrity* & 0.615\\
        DEP* & 0.527\\
        Data Execution Prevention & 0.615\\
        MPK* & 0.527\\
        Memory Protection Keys & 0.527\\
        MPX & 0.527\\
        Memory Protection Extensions* & 0.615\\
        ROP & 0.416\\
        Return Oriented Programming & 0.000\\
        SGX** & 0.245\\
        Software Guard Extensions & 0.288\\
        Stack Canaries & 0.464\\
        SFI & 0.527\\
        Software Fault Isolation & 0.464\\
        SDSec & 0.464\\
        \hline
    \end{tabular}
    \label{TFIDF_Security_Values}
    \end{table}

To address \textbf{RQ1}, the security terms that appear the most frequently with a TF-IDF value of 0.3 or smaller are ASLR and SGX as seen in Figure \ref{fig:security_TFIDF}, and are denoted by two stars (**) in Table \ref{TFIDF_Security_Values}.

\begin{figure}
    \centering
    \includegraphics[scale=.25]{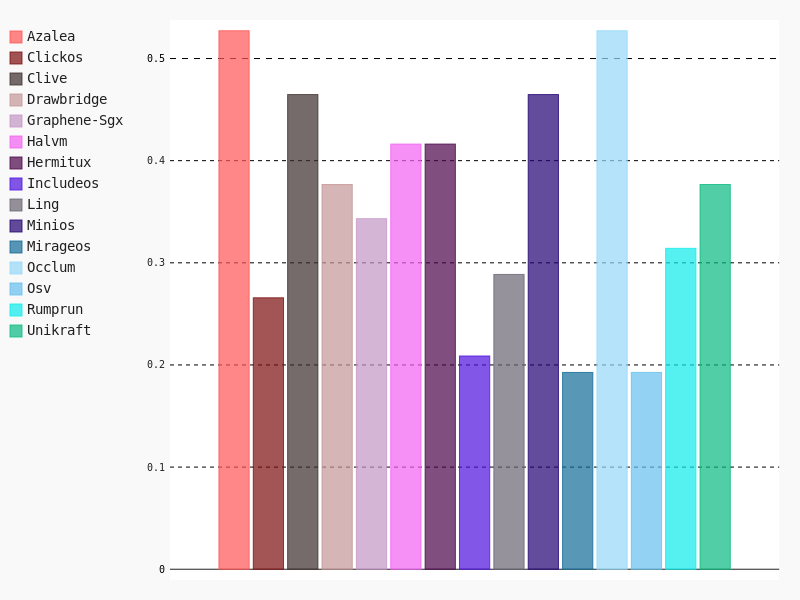}
    \caption{Unikernel TF-IDF Values}
    \label{fig:unikernel_TFIDF}
\end{figure}

The unikernel term TF-IDF values can be seen in Figure \ref{fig:unikernel_TFIDF} and Table \ref{TFIDF_Unikernel_Values}. The term in the left column of the table is the unikernel name, with its associated TF-IDF value in the right column. The TF-IDF values for unikernel terms are best interpreted to identify the most frequently discussed unikernels. These unikernels with a TF-IDF value of 0.4 or less are, IncludeOS, MirageOS, OSv, Rumprun, Graphene-SGX, and Ling and are denoted by a single star (*) in Table \ref{TFIDF_Unikernel_Values}. Missing unikernel terms are Nanos, Unik, Guk, Runtime.js, and Torokernel.

\begin{table}[htbp]
    \centering
    \caption{Unikernel TF-IDF Values}
    \begin{tabular}{|l|c|c|}
        \hline
        Term & TF-IDF Value\\
        \hline
        Azalea & 0.527 \\
        Clickos & 0.265 \\
        Clive & 0.464 \\
        Drawbridge & 0.376 \\
        Graphene-Sgx* & 0.343 \\
        Halvm & 0.416 \\
        Hermitux & 0.416 \\
        Includeos* & 0.208 \\
        Ling* & 0.288 \\
        Minios & 0.464 \\
        Mirageos* & 0.192 \\
        Occlum & 0.527 \\
        Osv* & 0.192 \\
        Rumprun* & 0.314 \\
        Unikraft & 0.376 \\
        \hline
    \end{tabular}
    \label{TFIDF_Unikernel_Values}
    \end{table}

\section{Discussion} \label{Discussion}

The TF-IDF values for the terms ASLR (0.288) and SGX (0.245) indicate that they are common topics in the corpus. Manual inspection of the occurrences for ASLR reveals a variety of uses. In \cite{unikernelIsolation_MPK} the term ASLR appears in the context of bypassing ASLR with a vulnerability in order to retrieve sensitive kernel data. However, the encasing paragraph is discussing how the proposed security technique (in this paper MPK) provides protection \emph{despite} the vulnerability. In contrast, \cite{UnikernelSecurityPerspective} directly addresses the lack of ASLR. This manual process reveals that while a TF-IDF value will indicate term frequency, it does not provide any context for the frequency. Addressing these variations through context utilization will be discussed in Section \ref{FutureWork}.

It is also more difficult to account for the context of security terms in regard to specific unikernels. While some literature focused on creating a new unikernel \cite{Uniguard}, other literature investigate multiple unikernels \cite{FreshLookAtIsolationPlatforms}, and still other literature choose one unikernel to make generalizations about \cite{michaels2019assessing, UnikernelSecurityPerspective}. In the first and last case, it is possible to identify specific unikernel(s) and then correlate their related security terms appropriately. However, the second case would be difficult to fully automate and guarantee accurate results.

A side effect of accounting for different variations of terms is reflected in the results: acronyms are counted separate from their expanded form. While the term ``Control Flow Integrity'' doesn't appear in many papers, its acronym ``CFI'' does, resulting in a variable value for, effectively, the same term. This is discussed further in Section \ref{FutureWork}.

These points all relate to answering \textbf{RQ4}. Where quantitative results provide answers to \textbf{RQ1}, \textbf{RQ2}, and \textbf{RQ3}, and can be used to provide some insight into unikernel security, relying solely on those values does not provide a complete picture. As the discussion in this section indicates, more specialized quantitative methods are needed to provide an accurate answer for \textbf{RQ4}. This is also discussed further in Section \ref{FutureWork}.

While creating the artifact and evaluating preliminary results, several challenges were discovered. Many papers include a ``keyword'' section which will affect the term frequency calculation, either by including or omitting words. The impact this has on the value is inappropriate because the term is not used in any context related to the paper, but rather metadata to assist external entities (researchers, search engines, etc.) While manual inspection and removal of these, and other, sentences from the artifact is possible, it would be infeasible to do for a large corpus. Additionally, not all security considerations can be clearly identified through acronyms or well known terms such as ASLR or DEP. For example, privilege escalation can be described as ``elevating to root'', ``gaining root permissions'', ``getting administrative access'', or ``getting admin credentials''; generating an exhaustive list of all such phrases would not be feasible.

The Python3 package PyPDF2 also revealed some challenges in the parsing of PDF documents. Variations in the way PDF documents are created result in inconsistencies in PyPDF2's output, one example being spacing between words. In examining this problem, the majority of sentences did not contain any spaces, but seemingly without rational a few sentences would. This complicates the construction of the regex and sentence validation, as the arrangement of acronyms cannot be guaranteed to be unique across word boundaries. In this research only one document\cite{michaels2019assessing} was discovered where this occurred, however as additional papers are added they should be monitored for this behavior.

The TF-IDF values indicate that some security terms are more prevalent in unikernel papers than expected. To provide better context the artifact could be improved to correlate unikernels and their related security terms. This and other possibilities are discussed in Section \ref{FutureWork}.

\section{Related Work} \label{RelatedWork}

TF-IDF has been used as a comparison metric for other term frequency calculation algorithms \cite{TrendingTerms, tfidfTermWeights}. 

In order to get the most accurate calculation, TF-IDF is best suited to parsing documents containing a single topic. \cite{TrendingTerms} proposed a different method which utilized a weighted log-odds ratio with an informative prior to calculate term frequency given a time range. The research used this method to calculate frequency with ``noisy'' text extracted from cybercrime forums.

\cite{tfidfTermWeights} calculated the significance of a term ``locally'' then used that value to help determine the terms ``global'' significance, utilizing TF-IDF as the catalyst.
\begin{quote}
``This novel perspective is a new avenue to develop more novel retrieval models, and it extends the original TF-IDF term weights to model microscopic phenomena at the document-location level, rather than macroscopic phenomena at the document level.''
\end{quote}

While there is no directly relatable work calculating term frequency for unikernels, TF-IDF is strongly tied to machine learning, natural language processing, and document typing \cite{textMiningTF_IDF, DocumentClustering} which have all become very popular. 

\section{Future Work} \label{FutureWork}

There are several expansions to the artifact that could improve the results. 

All variations of a security term and its acronym could be combined into one value, giving better insight into its frequency. This could be accomplished by adding a numerical field to the term list. After the individual values are determined, those that share a common numerical value would be summed, and then the new sum used in the TF calculation.

To evaluate the frequency of security terms in relation to specific unikernels, a two-fold approach could be used. One simple case exists if a paper is discussing a single unikernel. Any security term(s) discussed would be done in the context of that unikernel. A more complex case involves discussion of multiple unikernels throughout the research paper. The sentences discussing the security term(s), and possibly those sentences surrounding it, would need closer inspection to identify the context. This could be done by expanding the regex search for unikernel terms to include one or two sentences before or after a security term. If even more context were needed, such as determining a positive or negative connotation to the context, it could transform into a natural language processing (NLP) problem.

The corpus of papers vary greatly in size and therefore word count (one is over one hundred pages \cite{michaels2019assessing} and another only twelve \cite{Unikernel:LibOSForTheCloud}) an alternative algorithm could be used that accounts for these variations. 

A future objective is to correlate the occurrence of security terms to their corresponding unikernel. This component could provide better insights into the state of unikernel security. The research presented here only shows the frequency of terms, not how they relate. By adding another structure to track security terms as they appear in context to unikernels, a better understanding of unikernel security topics could be evaluated. This could be done in a manner similar to that described above for frequency calculation of security terms to specific unikernels.

Artifact improvements are not the only potential research avenue. A more dynamic classification guide for terms and phrases could expand security insights. A particular problem that arises is the creation of new acronyms based off existing ones. For instance MMDSFI, which stands for MPX-Based (Memory Protection Extensions), Multi-Domain SFI (Software Fault Isolation)\cite{Occlum} would trigger MPX and SFI, but not the actual acronym used. Not only would the count be inaccurate for MPX and SFI, but the current artifact would never detect nor count MMDSFI. These specialized terms would likely only appear in the originating document, but could create bias due to their unique nature. A more dynamic approach could possibly detect these terms and expand the initial list. This would once again, likely transform the work into an NLP problem.

\section{Conclusion} \label{Conclusion}

This research utilized a literature review methodology to collect a corpus of 33 unikernel research papers, which revealed a downward trend in unikernel research beginning in 2019 and an overall theme of unikernel development instead of security development. TF-IDF frequency analysis of security terms in the corpus reveals gaps that indicate potential vulnerabilities may be reintroduced with unikernels' reduced capabilities. Terms for foundational security mitigations like Memory Protection Extensions, Data Execution Prevention, and Control-Flow Integrity were amongst the least frequently occurring despite their widespread adoption for hardening systems against common attacks.

While development simplicity and minimalism are touted as unikernel benefits, the lack of discussion around impacts of forfeited defenses hints at an underestimation of the risks. Quantitative analysis of the literature demonstrates not just the ability to identify publication trends, but may also reveal assumptions and detect oversights in niche security domains. Unikernels show promise for improving cloud infrastructure security by shrinking the attack surface, but more research is needed to explore trade-offs and develop multilayered protections tailored for these special-purpose virtual machines. The gaps highlighted in this study can help guide future research efforts toward securing unikernels against attacks without sacrificing their performance advantages. Furthermore, this research demonstrates a literature review and TF-IDF methodology which should be widely applicable to identify trends and provide insights into niche security domains.

\balance
\printbibliography


\end{document}